\documentclass[preprint,showpacs,preprintnumbers,amsmath,amssymb]{revtex4}

\usepackage{graphicx}
\usepackage{dcolumn}
\usepackage{bm}

\begin{document}

\preprint{APS/123-QED}

\title{Surface Geometric and Electronic Structure of BaFe$_2$As$_2$(001)}

\author{V.B. Nascimento$^{1,*}$, Ang Li$^2$, Dilushan R. Jayasundara$^2$, Yi Xuan$^2$, Jared O'Neal$^2$, Shuheng  Pan$^2$, T. Y. Chien$^3$, Biao Hu$^3$, X.B. He$^1$, Guorong Li$^3$, A. S. Sefat$^4$, M. A. McGuire$^4$, B. C. Sales$^4$, D. Mandrus$^4$, M.H. Pan$^4$, Jiandi Zhang$^1$, R. Jin$^1$, and E.W. Plummer$^1$}


\affiliation{$^1$Department of Physics and Astronomy, Louisiana State University, Baton Rouge, LA 70803-4001}
\affiliation{$^2$Department of Physics and Texas Center for Superconductivity, University of Houston, Houston, Texas 77204-5002}
\affiliation{$^3$Department of Physics and Astronomy, The University of Tennessee, Knoxville, TN 37996}
\affiliation{$^4$Oak Ridge National Laboratory, Oak Ridge, TN 37831}

\date{\today}

\begin{abstract}
BaFe$_2$As$_2$ exhibits properties characteristic of the parent compounds of the newly discovered iron (Fe)-based  high-{\it T}$_C$   superconductors. By combining the real space imaging of scanning tunneling microscopy/spectroscopy (STM/S) with momentum space quantitative Low Energy Electron Diffraction (LEED) we have identified the surface plane of  cleaved BaFe$_2$As$_2$  crystals as the As
terminated Fe-As layer - the plane where superconductivity occurs. LEED and STM/S data on the BaFe$_2$As$_2$(001) surface indicate an ordered arsenic  (As)  - terminated metallic surface without reconstruction or lattice distortion. It is surprising that the STM images the different Fe-As
orbitals associated with the orthorhombic structure, not the As atoms in the surface plane.
\end{abstract}

\maketitle

The recently discovered superconductivity in Fe - based compounds marked an important advance in the endeavor to understand the mechanism for high transition temperature ({\it T}$_C$) superconductivity \cite{ref1,ref2}.  To date, four homologous series of Fe - based superconductors have been identified, commonly denoted as 1111 phase (REFeAsO or REFePO with RE = rare earth), 122 phase (AEFe$_2$As$_2$ with AE = alkaline earth, or (Sr$_4$Sc$_2$O$_6$)Fe$_2$P$_2$ \cite{ref3}), 111 phase (AFeAs with A = alkali metal) and 11 phase (FeTe or FeSe).  While superconductivity can be achieved by either chemical doping or applying external pressure, the undoped parent compounds exhibit rich structural and physical properties. Of particular importance is that all of the parent compounds undergo two successive phase transitions, a structural transition from a high symmetry tetragonal to a low symmetry orthorhombic phase, and a magnetic transition driven by antiferromagnetic (AFM) order \cite{ref4,ref5,ref6}.   Superconductivity  occurs when both of these transitions are suppressed \cite{ref7,ref8}.

Among the four homologous series, the 122 system is unique.  In this system superconductivity can occur without completely suppressing the AFM order --- there is a noticeable region in the phase diagram where superconductivity and AFM coexist \cite{ref11,ref12}.  Thus, this is an ideal system for studying the interplay between superconductivity and magnetism.  Single crystals of  these new 122 materials, which form layered structures,  can be easily cleaved to expose large flat surfaces. It is important both intellectually and technologically to address the physical properties of the surface of an ordered bulk material (single crystal), resulting from the effects of broken symmetry. Our experience with complex transition-metal oxides has been that small structural distortions at the surface can have quite dramatic consequence on the physical and chemical properties \cite{ref13,ref14}. A practical motivation for this work is related to the fact that several  powerful experimental techniques, such as Angle Resolved Photoemission Spectroscopy (ARPES) and Scanning Tunneling Microscopy/Spectroscopy (STM/S), are by their very nature surface-sensitive techniques.  To interpret the experimental data requires a thorough understanding of the surface, both the structural and electronic properties.  New states induced by creating a surface, such as surface states and/or states associated with surface reconstruction, may obstruct the characterization of bulk properties with STM/S and ARPES.

In this Letter, we report results obtained from the (001) surface of BaFe$_2$As$_2$ single crystals  using both STM/S and LEED techniques.  The BaFe$_2$As$_2$ compound forms a tetragonal structure with the {\it I4/mmm} space group symmetry at high temperatures ({\it T}), consisting of alternatively stacking Ba and Fe-As layers [see Fig. 1(a)] \cite{ref9,ref10}.  The structural transition, from high-{\it T} tetragonal ({\it I4/mmm}) to low-{\it T} orthorhombic ({\it Fmmm}) symmetry, occurs near 140 K, with corresponding unit cells schematically presented in Figs. 1(a) and  1(b) \cite{ref10}.  Furthermore, a collinear AFM ordering sets in around the same temperature, {\it i.e.} {\it T}$_N$ $\sim$ 140 K \cite{ref10}.

For the STM/S measurements, BaFe$_2$As$_2$ single crystals \cite{ref10} were first pre-cooled to $\sim$ 20 K in the UHV environment, then cleaved {\it in situ} to produce a clean (001) surface.  After cleavage, it was immediately inserted into the STM head, which was already at the base temperature ($\sim$ 4.3 K).  The subsequent STM/S measurements were all performed at 4.3 K.  For LEED measurements,  BaFe$_2$As$_2$ single crystals were  cleaved {\it in situ} under a base pressure of 2 $\times$ 10$^{-9}$ Torr at room temperature, producing a mirror-like (001) surface. After cleavage, the sample was immediately introduced into the LEED chamber maintained with a base pressure of 7.0 $\times$ 10$^{-11}$ Torr, and cooled down to 20 K for experimental data collection.

Figure 2(a) shows an 800 \AA \ $\times$ 800 \AA \ topographic image of a cleaved BaFe$_2$As$_2$ surface taken at 4.3 K in constant current mode. It reveals an atomically flat surface with many atoms (white dots) scattered, clustered, or piled on the surface without forming any ordered array.  These atoms on the surface are relatively easy to move by the tip when the tip is very close to the surface.  In addition to the atoms scattered on the surface, there are other features shown as dark spots and very weak chain-like structures in the atomically flat surface plane.  Figure 2(b) is the constant-current topographic image when zooming into the area indicated by the white square box in Fig. 2(a).    By pushing the tip closer to the surface (lowering the junction resistance), more details are revealed: (1) A ``square''-like lattice with the in-plane lattice constant of $\sim$ 5.6 \AA \ appears clearly. Such a large lattice constant reflects a termination of primary {\it p}(1$\times$1) structure truncated from an orthorhombic rather than a tetragonal bulk crystal (see Fig. 1); (2) The larger white dots are the scattered Ba atoms without ordered structure; (3) Some dislocations or structural boundaries, shown as smaller white dots, form chain-like structures; and  (4) Details of the darker spots and the zipper-like dark features are seen.

A line profile of the image at the location indicated by the white line in Fig. 2(b) is presented as an inset.  It illustrates that the corrugations of ``square''-like lattice are extremely small, with only 0.02 \AA. It is worth  emphasizing that the image of STM reflects the morphology profile of surface density of states (DOS) instead of lattice of atoms.  For example, Fig. 1(b) shows that there are two enviromentally distinct As atoms in each {\it p}(1$\times$1) orthorhombic surface unit cell (5.6 \AA$\times$5.6 \AA), because of the motion of the Fe atoms in the second layer, but  the STM image only shows one object per unit cell.  The inset of Fig. 2 (c) schematically presents the two distinct As atoms in the surface unit cell, one with four Fe ions slightly close to it while the other one with the Fe atoms slight away from it, due to the orthorhombic distortion.  STM measurements imply the existence of  different Fe-As orbitals associated with the orthorhombic structure, since only the DOS of one type of As atoms was detected.
By careful examination of the ``square''-like lattice, one can observe the small orthorhombicity distortion ($a' \neq b'$).   This is qualitatively consistent with the surface structure of the orthorhombic bulk at low temperatures.   The unit cell probed by STM reveals exactly the {\it p}($1\times1$) unit cell as terminated from the bulk orthorhombic structure (or similar to a $(\sqrt{2} \times \sqrt{2})R45^{o}$ one from the termination of a tetragonal structure). The current image shown in Fig. 2(c), obtained simultaneously with the constant current topographic image shown in Fig. 2(b), is the map of the tunneling current. While the topographic image in Fig. 2(b) was taken in the constant current mode, the image in Fig. 2(c) reflects the very fine feedback signal as the tip is scanned along the surface. This current image is known to remove the large corrugation of features and reveals the details which would otherwise be overwhelmed by the large background features. Figure 2(c) shows that the dark features in Fig. 2(b) are not caused by missing atoms of the  lattice, but are due to the slightly lower integrated DOS in these areas. These features have been confirmed repeatedly on all BaFe$_2$As$_2$ samples we have studied.

Figure 2(d) shows a tunneling spectrum taken at 4.3 K on the flat area with {\it p}($1\times1$) lattice and absence of either adatoms or defects.  This spectrum is representative to all locations as long as the tip is not located on top of or close to the bright scattered atoms. Even on top of the dislocations and in those dark features as appearing in Fig 2(b), the tunneling spectra still preserve the same line shape, with slightly lower total intensity. The spectrum is asymmetric about the Fermi energy (V = 0).  However, there are noticeable density of states (DOS) features at - 55mV and + 45 mV [indicated with arrows in Fig.2(d)], while their origins are yet to be identified.

From the crystal structure point of view, there are only two possibilities for cleavage: one is between Ba and As planes (see Fig. 1), and the other is between As and Fe planes.  The latter is very unlikely because of the strong bonding between As and Fe.  In the former case, it is also expected to see some differences between the two different surfaces: the As termination and the Ba termination (Fig. 1), which is also never observed. After  experiments on many crystal samples, we are convinced that the cleaved surface of BaFe$_2$As$_2$ is As terminated and the Ba layer in the bulk is destroyed by cleaving.  The random coverage of the scattered and clustered atoms we see in STM images (white dots) are the Ba atoms remaining on the surface after cleaving.  Obviously, the Ba atoms do not maintain their original lattice structure nor do they form any new superstructure.  The observed ``square"-like lattice is the As-terminated {\it p}($1\times1$) surface of the low-T orthorhombic phase [see Fig. 1(b)].  There is no evidence for surface reconstruction.

For a quantitative determination of surface termination and structure, LEED $I (V_{i})$ was  used.   Due to its low Debye temperature  \cite{ref9} the sample was cooled to low-T for good quality LEED $I (V_{i})$  data collection.  Shown in Fig. 3(a) is the LEED image taken at 20 K using a beam energy of 122 eV.   The unit cell derived from the LEED pattern appears to reflect a  (1$\times$1) surface for a tetragonal structure, rather than a surface for an orthorhombic structure as observed by STM. The beams that appear only in the orthorhombic phase seem to be missing, such as the (1,0) beam [expected at the position indicated by the arrow in Fig. 3 (a)].  The question is: are they completely missing or just weak, hidding in the intense diffuse background originating from the random Ba adatoms on the surface?

The advantage of our data collection procedure is that all of the data is stored allowing us to analyze carefully the intensities as a function of energy for any parallel momentum ($k_{//}$), as well as the energy dependent background. The black curve in Fig. 3(b) is the intensity as a function of beam energy (eV) for $k_{//}$ =(2,0), a strong diffraction spot in Fig. 3(a).  The red curve in Fig. 3(b) is the intensity for the orthorhombic {\it p}(1$\times$1) $k_{//}$=(1,0) direction. The energy dependence of the background is also presented. The blue curve at the bottom is the $k_{//}$=(1,0) intensity with the background subtracted.  As  can be seen, the (1,0) beam is very weak as expected from our calculations for the orthorhombic structure, but it is detectable with the careful analysis shown in Fig. 3(b). The theoretical $I (V_{i})$ curve for (1,0) is also presented for comparison. Some of the features (peaks) in (1,0) are in qualitative agreement with the calculations, but the poor signal to noise ratio and limited energy range of this beam make it not useful for determining the details of the surface structure. The low intensity of the beams associated with the orthorhombic structure can be attributed to two facts:  (1) The displacement of the Fe atoms due to orthorhombic distortions is in-plane and very small (~ 0.015 \AA) \cite{ref9,ref10};  (2) LEED is primarily sensitive to vertical displacement of the atoms.  Besides, the low Debye temperature of the constituting elements of BaFe$_2$As$_2$  also contribute to a small signal to background ratio.
As it can bee seen in Fig. 3(c), there is an intense energy dependent background associated with (2,0). For comparison, the intensity of a typical $I (V_{i})$  curve (and background) for another orthorhombic system, Sr$_3$Ru$_2$O$_7$(001) (our work), is also presented in Fig. 3(c). This bilayered perovskite is amenable to cleavage along (001), resulting in well ordered  Sr-O terminated surface.    The energy dependent background for Sr$_3$Ru$_2$O$_7$  is much less intense, thus giving a higher signal to background ratio. The large background for BaFe$_2$As$_2$(001) is a direct consequence of the Ba adatoms randomly scattered on the surface.

Given the fact that the surface has the structure with orthorhombic symmetry, the quantitative structural analysis was performed by constraining the surface structure to a {\it p2mm} symmetry (the 2D-symmetry representation of the {\it Fmmm} orthorhombic symmetry). Nine  symmetrically independent beams [(1,1), (2,0), (0,2), (2,2), (1,3), (3,1), (4,0), (0,4), (3,3) in orthorhombic notation], measured at 20 K over a total energy range of 2155 eV, were used in the structural analysis.  Shown in Fig. 4(a) are some of the {\it I ($V_i$ )} curves between 80 and 410 eV. The structural refinement has been performed through theoretical multiple scattering calculations described elsewhere \cite{ref15}, and the Pendry R-Factor ($R_P$ ) \cite{ref16} was used to quantify the theory-experiment agreement. 

Three different possible terminations for the BaFe$_2$As$_2$ (001) surface have been investigated:  (i) Ba-termination; (ii) As-termination; and 
(iii) Fe-termination in which the As-Fe-As trilayer is broken. The theory-experiment quantitative comparison demonstrates that the surface is As terminated, with $R_P$  = 0.24 compared to 0.57 for Ba termination and 0.45 for Fe termination.  The $R_P$  value obtained for the As terminated surface, (0.24 $\pm$ 0.03), indicates a very good theory-experiment agreement for such a complex system. The obtained results, as presented in Fig.  4(b) for the As - terminated model, confirm that the surface structure is basically orthorhombic bulk truncated. The only difference is that the refined thickness (2 $\times$ 1.37 \AA) of As-Fe-As trilayer is slightly larger than that (2 $\times$ 1.34 \AA) in the bulk, suggesting a possible small surface relaxation, but no reconstruction. The calculated beams of (1,1), (2,0), and (2,2) for the final As-terminated surface structure are displayed in Fig. 4 (a) to show the agreement with measured experimental results.

In summary, this work, combining real space imagining and momentum space diffraction, has shown that the cleaved surface of BaFe$_2$As$_2$  single crystals is an ordered As-terminated layer, whith disordered Ba-adatoms.
Quantitative analysis indicates that the surface structure represents the low temperature orthorhombic phase in bulk, without surface reconstruction. Our STS results show a finite density of states at Fermi energy  consistent with the metallic character in bulk, regardless the existence of randomly located Ba-adatoms. Importantly, our STM images reveal different Fe-As orbital structure associated with two distinct As atoms in the surface orthorhombic unit cell.

\begin{acknowledgments}
This effort was supported in part by Basic Energy Sciences, U.S. DOE, contract DE-AC05-000822725.  XBH, GL,TYC and EWP have received support from NSF and DOE (DMS\&E and NSF-DMR-0451163). JZ is supported by NSF under grant No. DMR-0346826. This work is also partly supported by the State of Texas through TcSUH and the Robert A. Welch Foundation. We would also like to thank DMSE and Division of User Facilities for work at CNMS.
\end{acknowledgments}


\begin{figure}
\includegraphics[width=0.65 \textwidth]{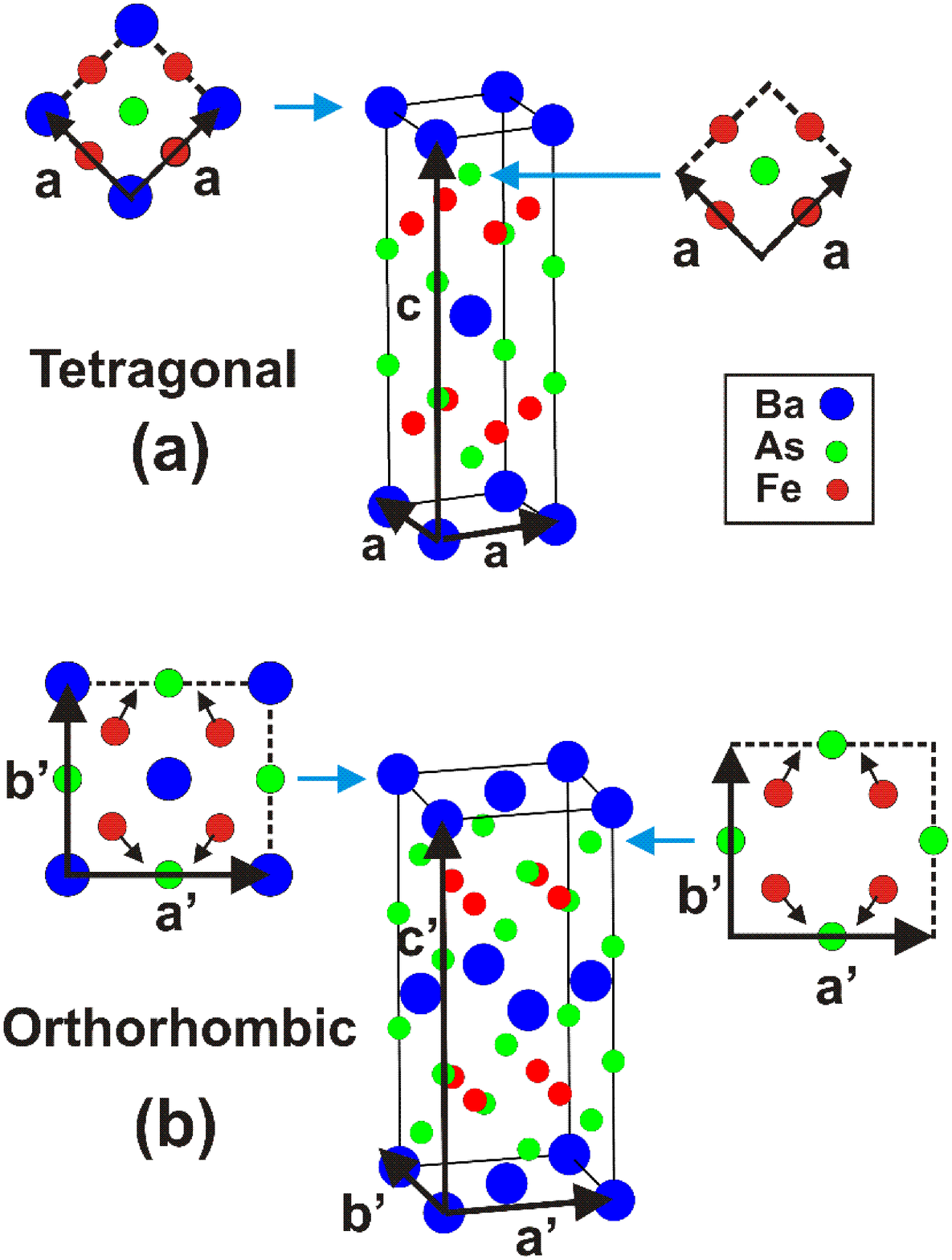}
\caption{\label{fig1} (a) BaFe$_2$As$_2$ bulk unit cell for the high temperature tetragonal phase.  The (001) surface unit cells are shown for both Ba (left) and As terminations (right).   (b) BaFe$_2$As$_2$ bulk unit cell for the low temperature orthorhombic phase. The (001) surface unit cells are presented for  Ba (left) and As terminations (right), where {\it a'} and {\it b'} lattice parameters are slightly different, i.e., 5.6146 \AA\ and 5.5742 \AA \, respectively. The    short black arrows show schematically the in-plane orthorhombic distortions of Fe atoms compared with the tetragonal phase. The Fe-Fe distances in the layer changes from 2.8020 \AA \ in the tetragonal phase, to 2.7870 \AA \ and 2.8070 \AA \ in the orthorhombic phase.}
\end{figure}

\begin{figure}
\includegraphics[width=0.65\textwidth]{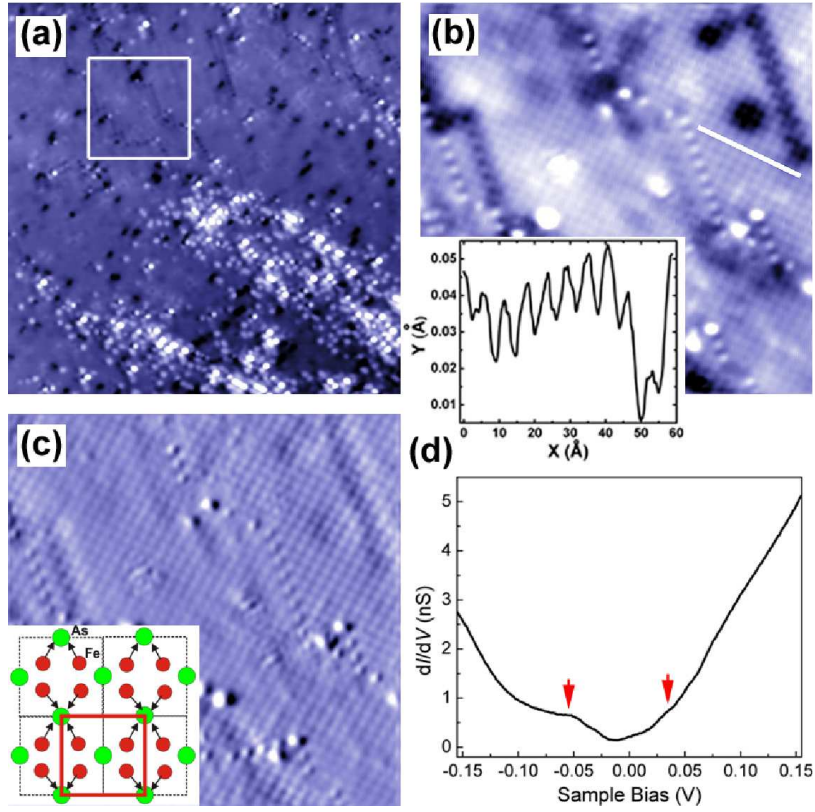}
\caption{\label{fig2} (a) 800 \AA $\times$  800 \AA \ constant current topographic image (V$_{sample}$ = -100 meV, I$_{tip}$ = 30 pA) taken on a cleaved surface at 4.3 K.   (b) 200 \AA $\times$ 200 \AA \ constant current topographic image (V$_{sample}$ = -20 meV, I$_{tip}$ = 8 nA) zoomed into the area indicated by the white square box in (a). A line-profile along the white line in (b) is presented in the inset.  (c) The current image obtained simultaneously with the image in (b). The inset schematically presents the two symmetrically distinct As atoms present in the orthorhombic  surface unit cell, which corresponds to the red square. (d) The representative tunneling spectrum taken on the surface in the areas far from those scattered atoms is presented.}
\end{figure}

\begin{figure}
\includegraphics[width=0.65\textwidth]{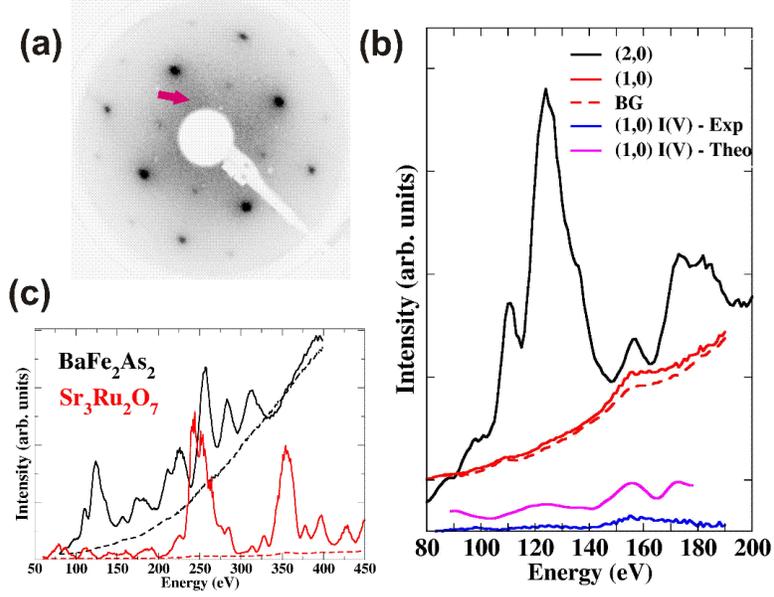}
\caption{\label{fig3}
 (a) LEED image for BaFe$_2$As$_2$(001) surface obtained for an energy of  122 eV and at a  temperature of 20 K. (b) Comparison between experimental $I(V_{i})$ curves for the (1,0) and (2,0) without background (BG) subtraction.  The dashed line represents the background  for (1,0).  The bottom curves are the theoretical and experimental  $I(V_{i})$ curves for (1,0). The theoretical intesities have been scaled for a better visualization. (c) Comparison between typical experimental $I(V_{i})$ curves of (2,0) beams for BaFe$_2$As$_2$ and Sr$_3$Ru$_2$O$_7$ orthorhombic systems, with corresponding backgrounds (dashed lines).}
\end{figure}

\begin{figure}
\includegraphics[width=0.65\textwidth]{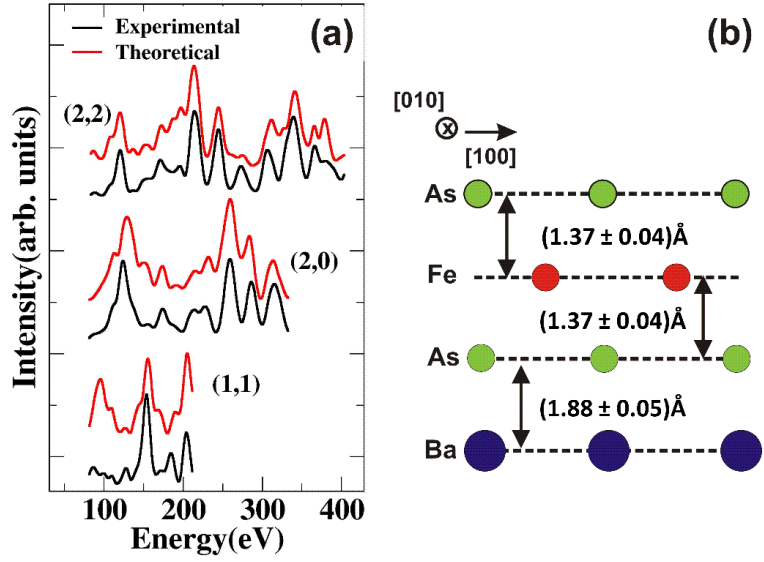}
\caption{\label{fig4}  (a) The measured $I(V_{i})$ curves of the diffracted beams (1,1), (2,0) and (2,2) compared with the theoretical ones for the final As-terminated surface structure. (b) The (001) surface structure obtained from LEED structural analysis. Bulk As-Fe and As-Ba interlayer distances are  1.3437 \AA \ and 1.8926 \AA \, respectively \cite{ref9}.}
\end{figure}

\end{document}